\documentclass[prx,letterpaper,nobalancelastpage,twocolumn,superscriptaddress,nofootinbib]{revtex4}

\usepackage{graphicx}
\usepackage{amsmath}
\usepackage{amssymb}
\usepackage[english]{babel}
\usepackage{color}
\usepackage[version=4]{mhchem}
\usepackage[hidelinks]{hyperref}
\usepackage{dsfont}

\usepackage{soul}
\newcommand{\ket}[1]{{| #1 \rangle}}
\newcommand{\bra}[1]{{\langle #1 |}}

\renewcommand{\t}[1]{\text{#1}}
\newcommand{\DtoP}[2]{
    { {}^3 \t{D}_{#1} \rightarrow {}^3 \t{P}_{#2} }
}
\newcommand{\PotoD}[2]{
    { {}^3 \t{P}_{#1} \leftrightarrow {}^3 \t{D}_{#2} }
}
\newcommand{\telecom}{\DtoP{2}{1}}

\newcommand{\UIUC}{Department of Physics, The University of Illinois at Urbana-Champaign, Urbana, IL 61801, USA}
\newcommand{\UC}{Pritzker School of Molecular Engineering, University of Chicago, Chicago, IL 60637, USA}

\renewcommand{\cite}[1]{\mbox{\citep{#1}}}

\addto\captionsenglish{}
\addto\captionsenglish{\renewcommand\thefigure{\arabic{figure}}}

\begin{document}

\title{Multiplexed telecom-band quantum networking with atom arrays in optical cavities}
\author{William Huie}
\affiliation{\UIUC}
\author{Shankar G. Menon}
\affiliation{\UC}
\author{Hannes Bernien}\email{bernien@uchicago.edu}
\affiliation{\UC}
\author{Jacob P. Covey}\email{jcovey@illinois.edu}
\affiliation{\UIUC}

\begin{abstract}
    The realization of a quantum network node of matter-based qubits compatible
    with telecom-band operation and large-scale quantum information processing
    is an outstanding challenge that has limited the potential of elementary
    quantum networks. We propose a platform for interfacing quantum processors
    comprising neutral atom arrays with telecom-band photons in a multiplexed
    network architecture. The use of a large atom array instead of a single atom
    mitigates the deleterious effects of two-way communication and improves the
    entanglement rate between two nodes by nearly two orders of magnitude.
    Further, this system simultaneously provides the ability to perform
    high-fidelity deterministic gates and readout within each node, opening the
    door to quantum repeater and purification protocols to enhance the length and
    fidelity of the network, respectively. Using intermediate nodes as quantum
    repeaters, we demonstrate the feasibility of entanglement distribution over
    $\approx 1500\,\t{km}$ based on realistic assumptions, providing a blueprint
    for a transcontinental network. Finally, we demonstrate that our platform
    can distribute $\gtrsim 25$ Bell pairs over metropolitan distances, which
    could serve as the backbone of a distributed fault-tolerant quantum
    computer.
\end{abstract}
\maketitle

\section{Introduction}\vspace{-3mm}
The development of a robust quantum
network~\cite{Cirac1997,Wehner2018,Kimble2008} will usher in an era of
cryptographically-secured communication~\cite{Pirandola2019}, distributed and
blind quantum computing~\cite{Jiang2007}, and sensor and clock networks
operating with precision at the fundamental limit~\cite{Komar2014}. Almost all
of these applications require network nodes that are capable of storing,
processing, and distributing quantum information and entanglement over large
distances~\cite{Kimble2008}. Nodes based on neutral atoms have the potential to
combine highly desirable features including minute-scale coherence and memory
times~\cite{Young2020}, scalability to hundreds of qubits per
node~\cite{Ebadi2020}, multi-qubit processing
capabilities~\cite{Saffman2010,Levine2019,Graham2019}, and efficient
light-matter interfaces at telecom wavelengths~\cite{Uphoff2016,Covey2019b, Menon2020}
based on optical cavities~\cite{Kimble2008,Reiserer2015}.

\begin{figure}[t!]
    \centering
    \includegraphics[width=8.6cm]{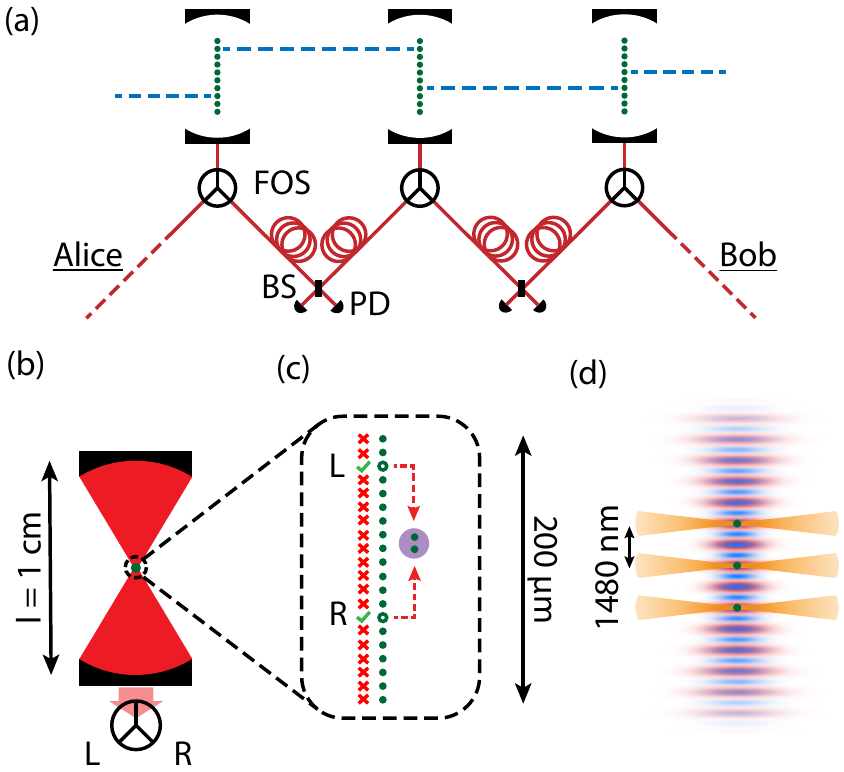}
    \caption{
        \textbf{Overview of the network architecture}. (a) Nodes based on arrays
        of atoms (green circles) in optical cavities generate a Bell pair over
        each link (blue dashes) to distribute entanglement between end-users
        Alice and Bob. We employ ``heralded'' entanglement generation based on
        photon interference on a 50:50 beamsplitter (BS). Fiber-optic switches
        (FOS) connect adjacent nodes at will by routing the photons from each
        cavity. (b) The near-concentric optical cavities have a mirror spacing
        of $\approx 1\,\t{cm}$ while the atom array spans a length of only
        $\approx 200$ $\mu$m. (c) The time signature of the photons on the
        detectors (PD) informs which atoms at each node are in a Bell state
        (green check marks). Subsequent, deterministic gates can be achieved by
        moving these atoms (dashed red arrows) and performing Rydberg entangling
        operations (purple circle). (d) A standing wave in the cavity traps
        atoms in a one-dimensional array (blue) to overlap with the highest
        field strength of the telecom mode (red).  Atoms are positioned with
        auxiliary optical tweezers (yellow) that also move the atoms.
    }
    \label{Fig1}
\end{figure}

Despite recent work establishing neutral atom-based
nodes~\cite{Ritter2012,Hofmann2012,Samutpraphoot2020,Langenfeld2021,Daiss2021,Dordevic2021},
a major bottleneck for the development of such a network is the exponential
attenuation and long transit time associated with sending single photons -- the
quantum bus that distributes entanglement -- throughout the
network~\cite{Reiserer2015}. Since the success probability per entanglement
generation attempt is low and success must be ``heralded'' via two-way
communication~\cite{Duan2001,Pfaff2013}, there is intense interest in developing
architectures that can ``multiplex'' many signals in parallel on each
attempt~\cite{Graham2013,Sinclair2014,Kaneda2015,Zhong2017,Wengerowsky2018}.
Multiplexing is necessary to construct networks much larger than the attenuation
length in optical fiber ($\approx$20 km in the telecom band~\cite{Corning2020}),
but it not sufficient. Intermediate ``repeater'' nodes are required to swap the
entanglement and teleport quantum information~\cite{Duan2001,Pirandola2019}.
Additionally, entanglement ``purification''
protocols~\cite{Dur2003,Bennett1996a,Kalb2017} are often needed to improve the
fidelity of the distributed quantum states.

\begin{figure*}[t!]
    \centering
    \includegraphics[width=13cm]{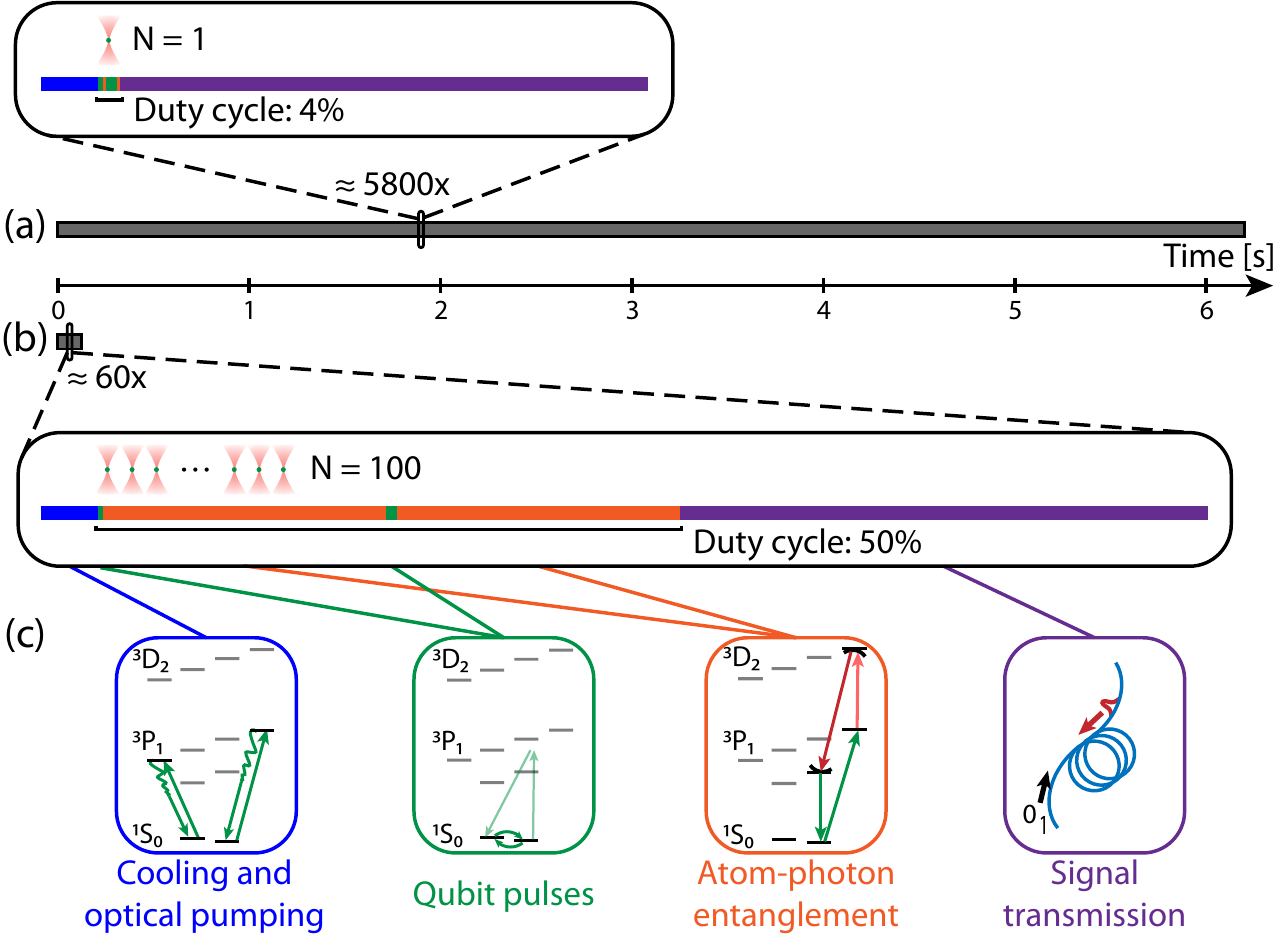}
    \caption{
        \textbf{Multiplexed remote entanglement generation over $L =
        100\,\t{km}$ with atom arrays}. (a) With only a single atom at each
        node, the low success probability necessitates an average of $\approx
        5800$ entanglement attempts. The solid gray bar shows this process with time moving
        to the right. The zoom shows a single attempt, in which the cooling and
        initialization of the atom (blue) and signal transmission over the link
        (purple) dominate the duration of each attempt. The duty cycle of
        entanglement-producing operations -- qubit rotations (green) and
        atom-photon entanglement (orange) [see Fig.~\ref{Fig3} for details] --
        is only $\approx 4$\%. (b) With an array of $N = 100$ atoms at each
        node, the success probability necessitates an average of only $\approx
        60$ multiplexed attempts; hence the overall time it takes to create
        entanglement (gray bar) is much shorter. Each attempt takes longer and
        has a much greater duty cycle of $\approx 50$\% for
        entanglement-producing operations. (c) A cartoon of the operations
        required for each attempt. The two qubit rotations and atom-photon
        entanglement operations follow the standard protocol for time-bin
        entanglement generation~\cite{Barrett2005,Bernien2013}.
    }
    \label{Fig2}
\end{figure*}

Here, we propose a quantum network and repeater node architecture that is
capable of high-rate, multiplexed entanglement generation, deterministic
inter-node quantum gates and Bell-state measurements for purification and
distribution of many-body states, while at the same time operating at telecom wavelengths where
low-loss optical fibers permit long-distance entanglement distribution. Our
architecture is based on arrays~\cite{Endres2016,Barredo2016} of individual
neutral ytterbium (Yb) atoms, an alkaline earth-like
species~\cite{Cooper2018,Norcia2018b,Saskin2019}, in large ($\approx
1\,\t{cm}$), near-concentric optical
cavities~\cite{Casabone2013,Nguyen2018,Davis2019,Zeiher2020} (see
Fig.~\ref{Fig1}). We consider a time-bin entanglement generation
protocol~\cite{Bernien2013} that combines a strong, $1.48$ $\mu$m-wavelength
transition~\cite{Covey2019b,Covey2019c} and long-lived nuclear spin-1/2 qubit
states of ${}^{171} \t{Yb}$ with temporal multiplexing along the array of atoms.

Based on recent progress with alkaline-earth atomic
arrays~\cite{Cooper2018,Norcia2018b,Saskin2019,Covey2019,Norcia2019,Madjarov2019,Madjarov2020,Wilson2019}
and realistic assumptions regarding the operation of these nodes, we show that
our multiplexing protocol can generate Bell pairs over $>1000$ kilometers
within the coherence time of the qubits, and is compatible with entanglement
purification protocols~\cite{Dur2003,Bennett1996a,Kalb2017} as well as the
distribution of many-body
states~\cite{Komar2014,Polzik2016,Bernien2017,Choi2021}. Our work lays the
foundation for a versatile metropolitan or transcontinental network through a
novel architecture that combines the use of Rydberg atom
arrays~\cite{Saffman2010,Browaeys2020}, cavity QED with strong atom-photon
coupling~\cite{McKeever2003,Birnbaum2005,Tiecke2014,Reiserer2015}, and
atom-array optical clocks~\cite{Madjarov2019,Norcia2019,Young2020} in one
platform for the first time.

\section{Multiplexed remote entanglement generation}
To motivate the proposed architecture, we begin with an overview of our multiplexed
time-bin networking protocol. Specifically, we consider the example of a network
link of length $L = 100\,\t{km}$. The associated two-way signal transmission
time per attempt is $\tau = 2 L / c$, where $c = c_0 / n$ is the speed of light
in optical fiber ($n = 1.4$) that includes both the quantum signal and classical
heralding signal; $\tau \approx 1\,\t{ms}$ for this distance. Per the methods
described below, we estimate that $\approx 5800$ entanglement attempts will be required if
there is only a single qubit (atom) at each node, resulting in a $\approx
0.16\,\t{Hz}$ entanglement generation rate. Figure~\ref{Fig2}(a) shows the full
process of successful entanglement generation with a zoomed view of each
attempt. The attempt time is dominated by signal transmission (see
Appendix~\ref{Sec:Dist} for full timing details) such that the duty cycle of
entanglement-producing operations is only $\approx 4\%$.

If instead we have $N=100$ qubits at each node and multiplex their signals as
described below, we can drastically decrease the number of required attempts to
only $\approx 60$ resulting in a $\approx50$-fold increase in the entanglement rate
to $8\,\t{Hz}$ at $L = 100\,\t{km}$. Figure~\ref{Fig2}(b) shows the full process
of successful entanglement generation for $N = 100$ atoms with a zoomed view of
each multiplexed attempt. In this case the duty cycle for entanglement-producing
operations is $\approx 50\%$. Although the time required per attempt is longer
when multiplexing across a large number of atoms, the favorable scaling in
success probability per attempt over long network links leads to substantially
improved entanglement generation rates compared to the case of a single atom.

\section{Description of the network architecture}
Before summarizing these results in more detail in section~\ref{sec:single-link}
and \ref{sec:network level}, we provide an overview of the atom array platform
and the atom-photon entanglement scheme. Further details on these topics can be
found in Appendices~\ref{Sec:Atomic} and~\ref{Sec:FWM}, respectively.

\subsection{Atom arrays in near-concentric optical cavities}
There has been intense interest in coupling neutral atoms to optical cavities
with small mode volumes such as
nanophotonic~\cite{Samutpraphoot2020,Dordevic2021,Menon2020} and fiber-gap
Fabry-P\'erot~\cite{Hunger2010,Haas2014,Brekenfeld2020} systems to enhance the
atom-photon coupling. However, these systems are not readily compatible with
large atom arrays (and single-atom control therein) due to their limited optical
accessibility. Additionally, the proximity of dielectric surfaces to the atoms
makes the prospect for robust, high-fidelity Rydberg-mediated gates uncertain as
stray electric fields limit the coherence of Rydberg
transitions~\cite{Sedlacek2016,Thiele2015}.

Meanwhile, near-concentric cavities with large mirror spacings ($\ell \gtrsim
1\,\t{cm}$) have recently been used with great success in myriad cavity QED
research directions~\cite{Casabone2013,Nguyen2018,Davis2019,Zeiher2020}, and
offer enough optical access to enable single-atom control in cavity-coupled atom
arrays. Crucially, the mirror spacing is similar to the size of glass cells used
in many recent high-fidelity Rydberg entanglement
studies~\cite{Levine2019,Graham2019,Madjarov2020,Wilson2019}. Further,
near-concentric cavities are widely used in trapped ion
systems~\cite{Casabone2013} that are also sensitive to transient electric fields from
dielectric surfaces~\cite{Teller2021}. Therefore it is reasonable to expect that
these cavities are compatible with deterministic Rydberg-mediated gates and Bell
state measurements needed in a quantum repeater and purification architecture.

We focus on a near-concentric system with $\ell = 0.975\,\t{cm}$ and radius of
curvature $R = 5\,\t{mm}$ for which the cavity stability parameter $\mathcal{G}
= 1 - \ell / R = -0.95$~\cite{Nguyen2018,Kawasaki2019}. We choose a single-sided
cavity, where the reflectivity of one mirror is much greater than the other to allow
photon passage, with a finesse of $50,000$. We couple this cavity to the
$\PotoD{1}{2}$ transition with wavelength $\lambda_\t{net} = 1480\,\t{nm}$ and
decay rate $\Gamma = 2 \pi \times 318\,\t{kHz}$. Based on these parameters, the
coupling strength to the cavity is $g_{34} \approx 2 \pi \times 1.53\,\t{MHz}$
and the single-atom cooperativity is $C \approx 16$ (for a detailed derivation
see Appendix~\ref{Sec:Atomic}).

We trap the atoms in a standing wave at $\lambda_\t{trap} = \lambda_\t{net} / 2
= 740\,\t{nm}$ to ensure maximal coupling with the telecom field (at
$\lambda_\t{net}$) in the cavity [see Fig.~\ref{Fig1}(d)]. The standing wave at
$\lambda_\t{trap}$ is fortuitously close to the `magic' wavelength for the
optical clock transition (${}^1 \t{S}_0 \leftrightarrow {}^3 \t{P}_0$) where the
two states have equal polarizability~\cite{Ye2008}: $\lambda_\t{m} =
760\,\t{nm}$. The expected $1 / e^2$ waist radius for this standing wave is
$w_\t{trap} \approx 14$ $\mu$m; the trap depth (and frequency) are free
parameters. Optical tweezers are employed to create an atom array from the
magneto-optical trap (MOT) before the standing wave is turned on, and the
tweezers are positioned to overlap the desired anti-nodes of the standing wave
[see Fig.~\ref{Fig1}(d)]. The standing wave provides strong axial confinement
with $\lambda_\t{trap} / 2$ spacing between the anti-nodes and guaranteed
maximal overlap with the anti-nodes of telecom cavity mode at $\lambda_\t{net}$,
and the tweezers provide strong transverse confinement.

\begin{figure}[t!]
    \centering
    \includegraphics[width=8.6cm]{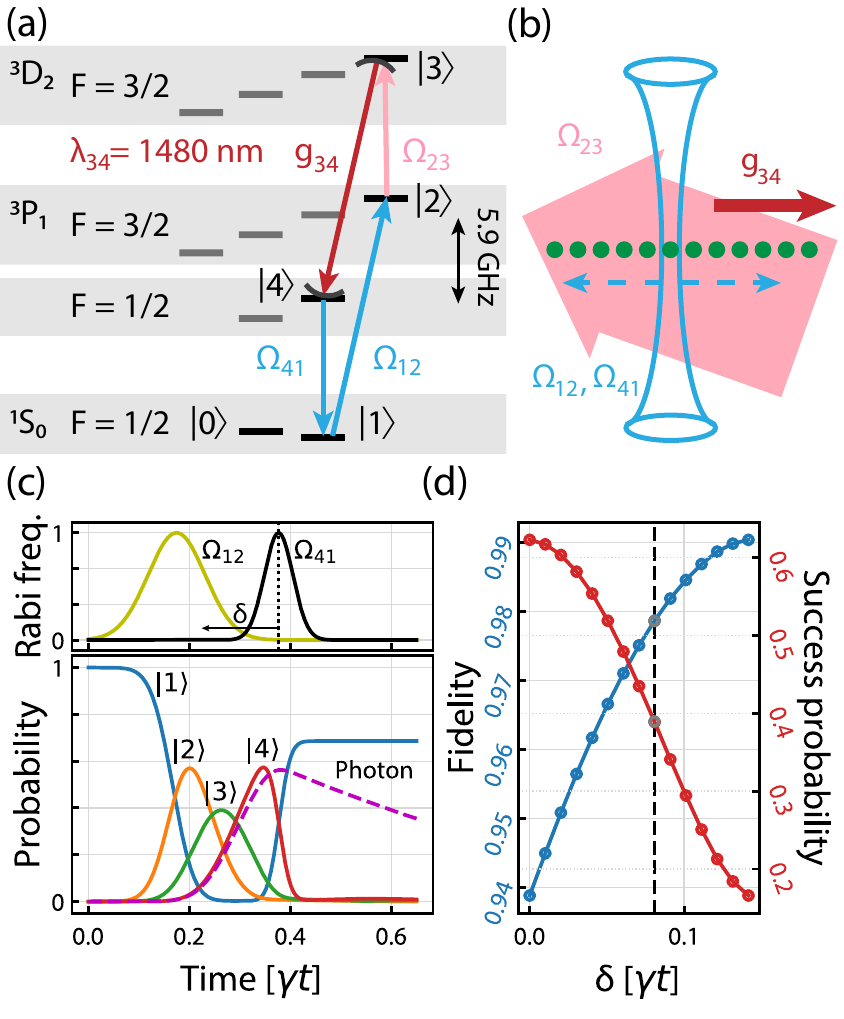}
    \caption{
        \textbf{Multiplexed remote entanglement via a four-pulse excitation
        scheme.} (a) A minimal diagram of the ${}^{171} \t{Yb}$ level structure
        showing two hyperfine Zeeman states in the ${}^3 \t{P}_1$ manifold as
        intermediaries. (b) Local application of $\Omega_{12}$ and $\Omega_{41}$
        on an atom-by-atom level is the primary mechanism for our time-based
        multiplexing scheme. (c) Analysis of the pulses and internal dynamics
        during the process as well as the temporal shape of the extracted photon
        that is entangled with the nuclear qubit in the ground state. The black
        arrow highlights that the relative timing of the two pulses
        $\Omega_{12}$ and $\Omega_{41}$ is a free parameter. The maximum Rabi
        frequencies of these pulses are $\{\Omega_{12}^\t{max},
        \Omega_{34}^\t{max}\}=\{13.2 \gamma, 23.0 \gamma\}$ and $\gamma = 2 \pi
        \times 180\,\t{kHz}$ is the decay rate of ${}^3 \t{P}_1$. (d) The
        resulting atom-photon entanglement fidelity and success probability vs
        the relative timing $\delta$ of $\Omega_{12}$ and $\Omega_{41}$ in units
        of $\gamma$. [(c) corresponds to $\delta = 0$.] We choose $\delta$ as
        shown in the black dashed line for the remainder of this work.
    }
    \label{Fig3}
\end{figure}

\subsection{Atom-photon entanglement via four-wave mixing}
We entangle the nuclear spin-$1/2$ qubit in the ground state of ${}^{171}
\t{Yb}$ with a $1480\,\t{nm}$-photon on the $\PotoD{1}{2}$ transition via a
four-pulse scheme that uses two Zeeman states within the ${}^3 \t{P}_1$ manifold
as intermediaries [see Fig.~\ref{Fig3}(a)]. The target state of our protocol is
the atom-photon Bell state $\ket{\psi}_\t{atom-photon} = (\ket{0}_a
\ket{\t{early}}_p + \ket{1}_a \ket{\t{late}}_p) / \sqrt{2}$, in which the atomic
qubit states \{$\ket{0}_a, \ket{1}_a$\} are entangled with the photon occupation
in an early and late emission time bin \{$\ket{\t{early}}_p,
\ket{\t{late}}_p$\}~\cite{Barrett2005,Bernien2013}. Such time-bin encoded states
are ideally suited for long-distance entanglement distribution via optical
fibers as they are robust against birefringence in fibers that would adversely
affect other encodings such as polarization encoded states. To create
$\ket{\psi}_\t{atom-photon}$ we start by preparing a superposition of the atomic
qubit states $(\ket{0}_a + \ket{1}_a)/ \sqrt{2}$. Then a coherent atomic pulse
sequence results in the emission of a photon into the cavity mode only if the
atom is in $\ket{1}$. The proposed four-level system that allows such a state
selective emission is shown in Fig.~\ref{Fig3}(a), and was inspired by similar
sequences that have recently been considered for alkali
species~\cite{Menon2020}. After the emission in the early time bin, a
$\pi-$pulse on the qubit states flips $\ket{0}_a$ and $\ket{1}_a$, and second
optical pulse sequence causes emission in the late time bin. This completes the
protocol and leaves the system in the target state $\ket{\psi}_\t{atom-photon}$.

We leverage the $F = 3/2$ and $F = 1/2$ hyperfine structure of the ${}^3
\t{P}_1$ manifold to provide the well-separated intermediate states $\ket{2}$
and $\ket{4}$, and we assume a magnetic field of $B\gtrsim100$ G although this
is not strictly necessary. We apply Gaussian pulses $\Omega_{12}$ and
$\Omega_{41}$ on a per-atom basis within the array [Fig.~\ref{Fig3}(b)] as the
primary mechanism for our time-based multiplexing scheme. $\Omega_{23}$ and
$g_{34}$ couple to all atoms globally, but are distantly off-resonant with
negligible differential effect on the qubit $\ket{0} - \ket{1}$ when
$\Omega_{12}$ and $\Omega_{41}$ are not applied to the atom. Hence, we raster
the tightly-focused $\Omega_{12}$ and $\Omega_{41}$ beams across the atoms such
that the position of the atom in the array is mapped to the time-stamp of the
photon emitted into the cavity.

We describe the optimization and analysis of the pulse design in
Appendix~\ref{Sec:FWM} and summarize our findings in Fig.~\ref{Fig3}(c). We
leave $\Omega_{23}$ at a constant value for the entire duration of the four-wave
mixing (FWM) protocol. We then transfer population from $\ket{1}$ to $\ket{2}$
with $\Omega_{12}$. These two fields populate $\ket{3}$, which is transferred to
$\ket{4}$ by the coherent cavity coupling $g_{34}$. Note that other schemes for
transferring population from $\ket{1}$ to $\ket{3}$, such as a two-photon
$\pi$-pulse detuned from the intermediate state $\ket{2}$, are expected to
further suppress double-excitation due to decay during the first half of the FWM
protocol to below 1\%. We then perform $\Omega_{41}$ to
coherently transfer the atomic population back to $\ket{1}$. The relative timing
of the $\Omega_{12}$ and $\Omega_{41}$ pulses introduces a trade-off between
process fidelity and success probability [Fig.~\ref{Fig3}(d)]. Essentially, the
process is limited by spontaneous emission from $\ket{4}$ which occurs at a rate
$\Gamma_{41} \approx 2 \pi \times 180\,\t{kHz} \approx g_{34} / 8.5$. Moving the
$\Omega_{41}$ pulse earlier mitigates the decay but reduces the probability of
success. Note that the remote entanglement scheme is heralded, so events that do
not produce photons only affect success rates, while events that produce photons
but leave the atom in the wrong state are classified as successful and lead to
infidelity.  We choose the values shown in Fig.~\ref{Fig3}(d) for which the
fidelity (success probability) of producing $\ket{\psi}_\t{atom-photon}$ with
the photon in the fiber is $\approx 0.98$ ($\approx 0.39$).

\section{Entanglement distribution across a single link}
\label{sec:single-link}
We now return to the discussion of entanglement distribution rates and we begin
by considering a single link between two nodes. Details of the analysis are
described in Appendix~\ref{Sec:Dist}. Figure~\ref{Fig4} shows the mean entanglement
rate in our multiplexed scheme versus the distance between the nodes for
different atom numbers $N$. For distances larger than $\approx 20\,\t{km}$, we
find a drastic improvement of the entanglement rate as more atoms per node are
used. At a distance of $100\,\t{km}$ we see a $\approx 50$-fold faster rate
when using $100$ atoms per node compared to the single-atom case (see also
Fig.~\ref{Fig2}). We find that the entanglement rate sees diminishing returns
for $N\gtrsim 200$ due to two main factors. First, the
probability of successfully creating a Bell pair asymptotically saturates at
$1$ such that larger numbers of atoms are not needed for suitably large rates.
Second, the time per entanglement attempt becomes dominated by the total time
required to perform the four-wave-mixing protocol for all the atoms at each
node, rather than the classical signal transmission time between them (see
Fig.~\ref{Fig2} and Appendix~\ref{Sec:Dist}). This second effect is clearly
visible at short distances below $\approx25$ km.

\begin{figure}[t!]
    \centering
    \includegraphics[width=8.6cm]{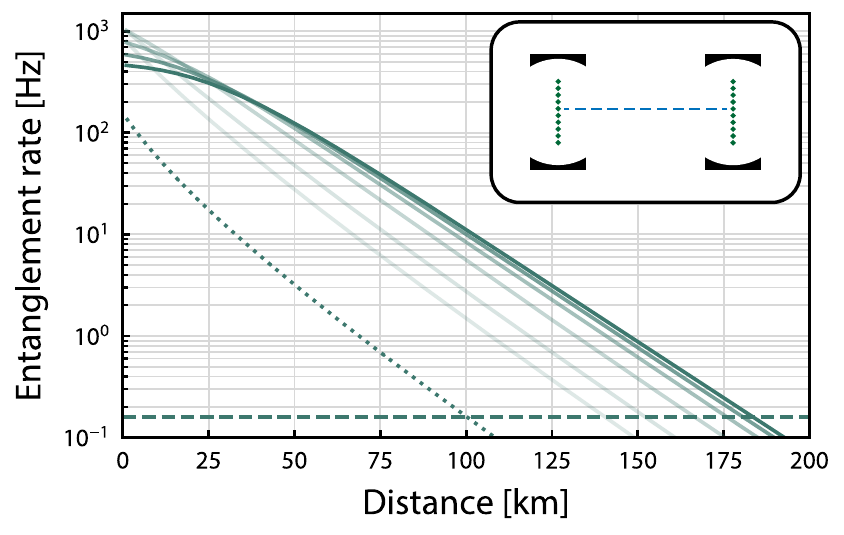}
    \caption{
        \textbf{Analysis of a single link}. (a) Multiplexed entanglement
        generation between two nodes, each containing an array of atoms in an
        optical cavity. The mean entanglement distribution rate versus the
        length of the link for various numbers of atoms $N$ is shown as an
        opacity scale for $N = \{ 10, 20, 50, 100, 150, 200 \}$ with $N = 200$
        being fully opaque. This scale is used in subsequent figures. The dotted
        line shows the entanglement distribution rate for $N = 1$. The
        horizontal dashed line shows a conservative estimate for the anticipated
        decoherence rate of the atoms. Here we focus on the rates associated with
        successfully generating a single Bell pair with one atom at each node
        (inset).
    }
    \label{Fig4}
\end{figure}

We compare the entanglement rate to the coherence time of the qubits in the
nodes. We assume a conservative lower bound of $T_2 = 1\,\t{s}$ for our
nuclear qubits, but note that it could approach the minute scale~\cite{Young2020}. Hence,
we consider distribution rates above $\Gamma_\t{coherence} = 1/(2 \pi T_2) =
0.16$ Hz to have a sufficiently high link efficiency~\cite{Hucul2015} for useful
entanglement. This criterion suggests that our platform will enable the
generation of entanglement over $\approx 180\,\t{km}$ using $N = 200$ atoms,
which is well within the reach of current technology~\cite{Ebadi2020}. For context, the current record for matter-based qubits is 1.3 km~\cite{Hansen2015}.

\section{Entanglement distribution using quantum repeater nodes}
\label{sec:network level}
We now turn to the use of intermediate repeater nodes to extend the range of
entanglement generation to greater distances. The goal is to connect these
intermediate links into a larger chain which we refer to as the
``network-level'' architecture. We break the length $L$ between end-users Alice
and Bob into $2^m$ segments with length $L_m = L / 2^m$, where $m$ is a
non-negative integer we call the ``nesting level'' of the network. 

\subsection{Overview of the protocol}
\begin{figure}[t!]
    \centering
    \includegraphics[width=8.6cm]{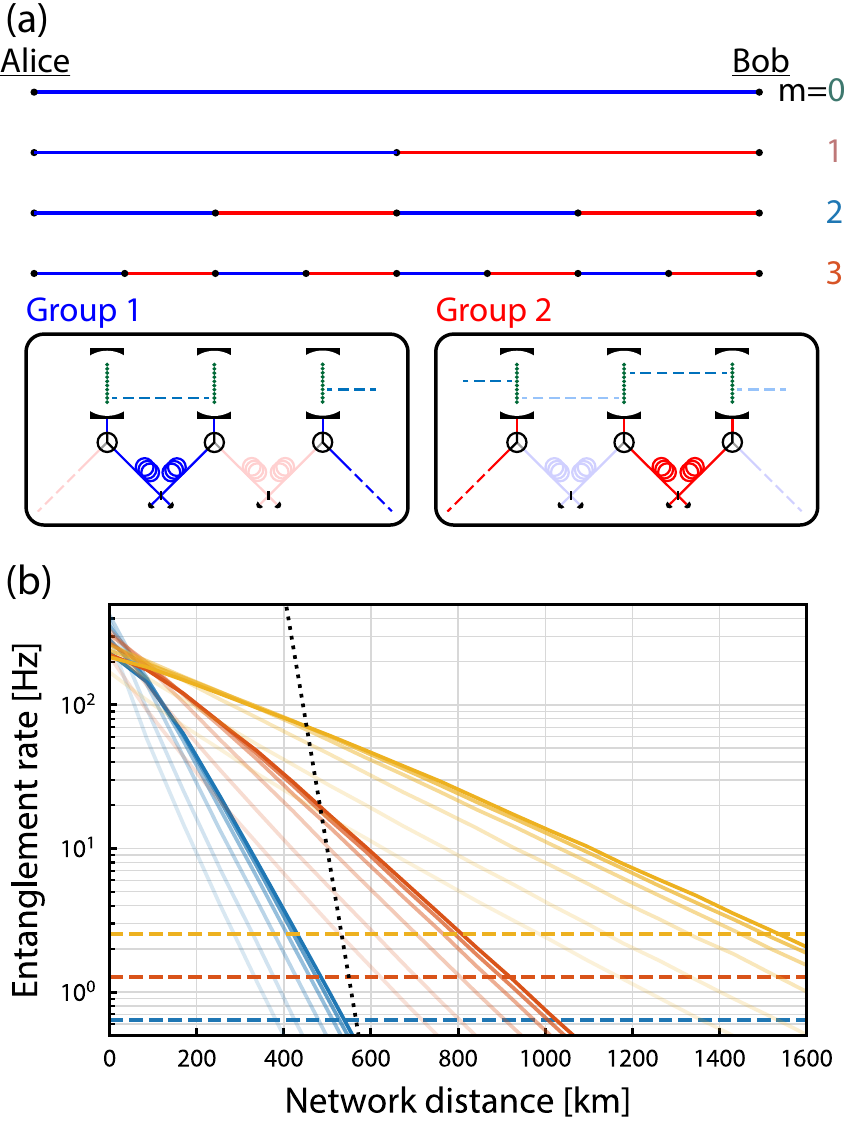}
    \caption{
        \textbf{Network-level entanglement generation}. (a) In order to 
        distribute entanglement to end-users Alice and Bob over greater
        distances, $2^m - 1$ intermediate repeater nodes are used, where $m$ is
        the nesting level. Bell pairs are generated in parallel within Group 1
        (blue) and Group 2 (red). Intermediate
        nodes have two atoms involved in Bell pairs. (b) Simulated entanglement
        distribution rates over the full network versus the network length $L$
        for nesting levels $m = 2$ (blue), $3$ (orange), and $4$ (yellow) with
        number of atoms per node $N$ shown as the same opacity scale as in
        Fig.~\ref{Fig4}. The dashed lines again show conservative estimates of
        the coherence of the qubits at each nesting level. Note that the number
        of qubits depends on $m$, so the estimated coherence is $2^m / (2 \pi
        T_2)$. The black dotted line shows for comparison the direct
        entanglement distribution rate by sending entangled photon pairs at a
        rate of $10\,\t{GHz}$~\cite{Pirandola2017}.
    }
    \label{Fig5}
\end{figure}

We divide the intermediate links into two groups in alternation such that
adjacent links are not in the same group [see Fig.~\ref{Fig5}(a)]. Our protocol
is based on the generation of Bell pairs across all Group 1 links in parallel
followed by all Group 2 links in parallel. Naively, the mean time required to
generate Bell pairs across all links is approximately twice the mean time
required for a single link. However, the number of attempts required to successfully create entanglement follows a geometric distribution and both groups must wait for the success of \textit{all} constituent links. Hence, we stochastically sample the distribution of
attempts for each link in both groups in order to estimate mean entanglement generation
rates at the network level (see Appendix~\ref{Sec:Dist} for details). Note that
if $N$ atoms are employed in the multiplexed entanglement generation in Group 1,
$N - 1$ atoms are available for generating entanglement in Group 2.

After the Bell pairs have been generated on Group 1 links, the constituent atom
at each node in these Bell pairs -- recognized by its time stamp -- must be
isolated and preserved from the subsequent operations on the Group 2 links. Our
protocol is based on transferring those qubits from the nuclear spin-$1/2$
ground state (${}^1\t{S}_0$) to an auxiliary computational
basis~\cite{Gorshkov2009} of the nuclear spin-$1/2$ metastable clock state
({}$^3\t{P}_0$) that has a lifetime of $\approx 20\,\t{s}$. Accordingly, we
leverage the (nearly-)clock-magic wavelength of the cavity standing wave-optical
tweezer trap system. The metastable clock state is transparent with respect to
the four-wave mixing sequence and a negligible relative phase is anticipated on
this auxiliary qubit. We expect that transferring the qubit to the auxiliary
basis will occur at a rate much faster than the entanglement generation rates
over distances of interest and therefore have a negligible effect on the total
rate. Rates of $\Omega_\t{clock} \approx 2 \pi \times 100\,\t{kHz}$ and a
transfer fidelity of $\gtrsim 0.99$ are anticipated with ${}^{171} \t{Yb}$~\cite{Young2020}.
Alternatively, the atom(s) could be moved away from the array and the laser
fields to preserve coherence during Group 2 operations.

With Bell pairs across all neighboring links, we now complete the end-to-end
entanglement protocol by entangling atomic pairs and performing deterministic
Bell-state measurements at each node to effectively reduce the nesting level of
the network by 1. Bell pairs between increasingly distant nodes are traced out
of the system through this process (see Appendix~\ref{Sec:RepPur}) until
end-users Alice and Bob directly share a Bell pair. We couple to highly-excited
Rydberg states to perform the required local deterministic
entanglement operations~\cite{Levine2019,Graham2019,Madjarov2020}, inspired by a recent
approach with alkaline-earth atoms coupling from the clock state to Rydberg
states in the ${}^3\t{S}_1$ series~\cite{Madjarov2020}. However, this
interaction occurs only over short distances, requiring the atomic pairs to be
re-positioned [see Fig.~\ref{Fig1}(c)]. The optical tweezers will remove the
atoms from the cavity standing wave and translate them to within several microns
of each other prior to Rydberg excitation. Tweezer-mediated coherent translation
of atomic qubits over such distances is routinely performed on the $\sim \t{ms}$
timescale with minimal
decoherence~\cite{Beugnon2007,Lengwenus2010,Schymik2020,Dordevic2021} and
Rydberg-mediated gates are on the $\lesssim\mu$s
timescale~\cite{Levine2019,Graham2019,Madjarov2020}. These steps are again much
faster than the entanglement distribution rates and are only performed when
remote Bell pairs have been successfully created, so we can neglect their effect
on the total rate. The expected near-term fidelity of Rydberg-mediated gates and
local measurement is $\gtrsim 0.99$~\cite{Covey2019,Madjarov2020}, which is high
compared to the fidelity of generating Bell pairs: $\approx0.98^2=0.96$ [see
Fig.~\ref{Fig3}(d)]. A detailed network fidelity budget is outside the scope of
this work.

\subsection{Summary of the results}
We consider the network-level entanglement distribution rate based on this
protocol for varied network length $L$, nesting level $m$, and atom number per
node $N$. We compare this rate against a conservative estimate of the coherence
of \textit{all} qubits in the system. Naturally, this depends on the nesting
level, and hence the network level coherence estimate is $\Gamma_\t{coherence}^m
= 2^m / (2 \pi T_2) = 0.16 \times 2^m\,\t{Hz}$. Figure~\ref{Fig5}(b) shows the
network level generation rate versus the network length for nesting levels $m
= 2,\, 3,\, 4$ with various atom numbers per node $N$ shown as an opacity scale.
We also compare against direct communication (without intermediate nodes) based
on entangled photon pairs at a wavelength of $1550\,\t{nm}$ with a repetition
rate of $10\,\t{GHz}$~\cite{Pirandola2017}. The direct communication rate falls
sharply, passing below our coherence time estimates at a distance of $\approx
600\,\t{km}$. We find that the achievable network length increases for higher
nesting level and saturates for $N \approx 200$ atoms. In particular, for $m =
4$ our system enables a network of $L \approx 1500\,\t{km}$.

\section{Multiple Bell pairs and entanglement purification}
\label{sec:multi-bell}
We now consider the generation of multiple Bell pairs with our system, which are
needed for more advanced protocols such as purification and logical encoding.
Entanglement purification (also known as
distillation)~\cite{Dur2003,Bennett1996a,Kalb2017} is based on taking two (or
more) Bell pairs and consuming them to generate a single Bell pair with higher
fidelity (See Appendix~\ref{Sec:RepPur}). Purification requires entanglement
operations between the local qubits in the pairs combined with single-qubit
readout within each node. The former will again be accomplished with
Rydberg-mediated gates~\cite{Levine2019,Graham2019,Madjarov2020} while the
latter will leverage the auxiliary qubit basis in the metastable clock state to
perform single-atom readout by scattering photons from the ${}^1 \t{S}_0
\leftrightarrow {}^3 \t{P}_1$ transition, to which the ${}^3 \t{P}_0$ clock
state is transparent~\cite{Monz2016,Erhard2021}.

\begin{figure}[t!]
    \centering
    \includegraphics[width=8.6cm]{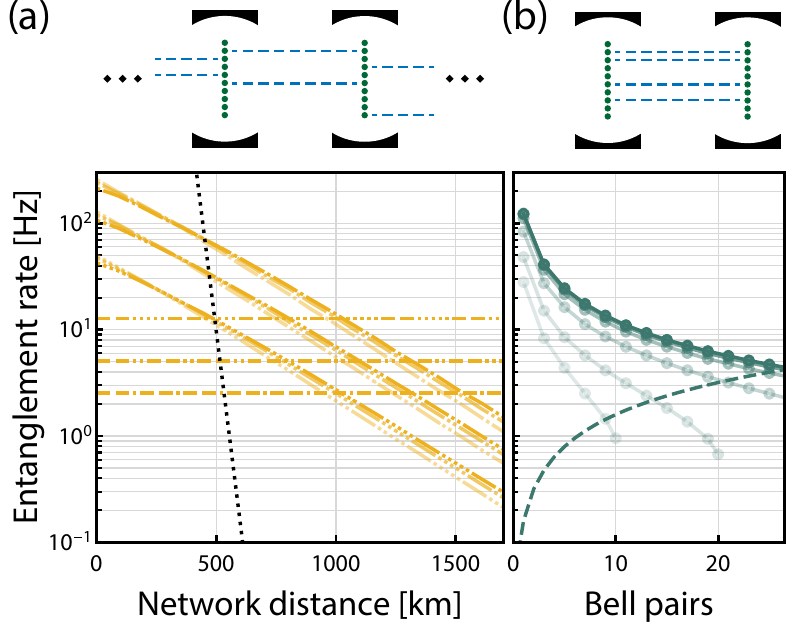}
    \caption{
        \textbf{Multiple Bell pairs at the network and single link levels}. (a)
        The simulated entanglement distribution rate for $m = 4$ versus network
        distance for one Bell pair (dot-dashed), two Bell pairs
        (dot-dot-dashed), and five Bell pairs (dot$^5$-dashed) with $N=\{ 100,
        150, 200 \}$ as an opacity scale. The black dotted line is again direct
        communication at $10\,\t{GHz}$. The horizontal dashed lines are the
        expected coherence associated with the total number of qubits. For
        multiple Bell pairs $B > 1$, this is $\Gamma_\t{coherence}^{m, B} = B
        \cdot 2^m / (2 \pi T_2)$. The maximum distance falls from $\approx 1500$
        to $\approx 1100\,\t{km}$ when increasing the number of Bell pair from
        one to two. (b) The entanglement distribution rate of a single link with
        distance $L = 50\,\t{km}$ to represent a metropolitan-scale network. The
        rate is plotted versus the number of Bell pairs with $N$ from 10 to 200 as an opacity
        scale. This shows a favorable scaling with $B$, and that $B = 26$ Bell
        pairs can be generated with $N = 200$, where the dashed line is again
        the expected coherence associated with the total number of qubits.
        Entanglement of $B > N$ pairs is impossible; hence rates for these data
        points are omitted.
    }
    \label{Fig6}
\end{figure}

To this end, we study the network-level entanglement generation rate versus
network length $L$ with $m = 4$ for various numbers of Bell pairs. We find that
rate associated with generating $B$ Bell pairs in a given attempt decays
exponentially with $B$; hence, we instead use a ``ladder" scheme analogous to
the network-level analysis. Specifically, we create $B$ Bell pairs one at a time
on each link [see Fig.~\ref{Fig6}(a)], and still divide the links into two
groups. Here again, we must sample the distribution of attempts before the
successful generation of each Bell pair on each link, and both Group 1 and 2 are limited by
the time for each constituent link to generate $B$ pairs.

We find that the simulated mean entanglement generation rate for $B = 2$ exceeds
the decoherence of the $B \cdot 2^m$ Bell pairs for distances up to $L \approx
1100\,\t{km}$.  These findings indicate that our platform may be compatible with
the development of a transcontinental terrestrial quantum network with
sufficiently high fidelity -- based on entanglement purification -- for
subsequent nontrivial operations. Interestingly, we find a favorable scaling
with $B$ and include $B = 5$ in Fig.~\ref{Fig6}(a), showing rates exceeding
decoherence for distances up to $L \approx 500\,\t{km}$.

Finally, we consider the possibility of generating many Bell pairs over a
metropolitan-scale link with $L = 50$ km for advanced error correction protocols or
for the distribution of many-body states such as logically-encoded
qubits~\cite{Fowler2012,Albert2020,Erhard2021}, atomic cluster or graph
states~\cite{Choi2019}, spin-squeezed
states~\cite{Polzik2016,Pezze2018,Pedrozo2021} or Greenberger-Horne-Zeilinger
(GHZ) states~\cite{Komar2014,Omran2019}. We analyze the entanglement
generation rate versus the number of Bell pairs per link for various $N$ in
Fig.~\ref{Fig6}(b). Crucially, we find that 26 Bell pairs can be generated for
$N = 200$ -- comparable with the largest GHZ states created locally to
date~\cite{Omran2019,Song2019,Marciniak2021} -- offering new opportunities for distributed
computing and error-corrected networking.

\section{Outlook and conclusion}
We have proposed a platform that combines the strengths of neutral atoms --
efficient light-matter
interfaces~\cite{McKeever2003,Birnbaum2005,Tiecke2014,Reiserer2015} with telecom operation~\cite{Uphoff2016,Covey2019b,Menon2020},
high-fidelity qubit operations and
measurement~\cite{Levine2019,Graham2019,Covey2019,Madjarov2020}, scalability to
many qubits~\cite{Bernien2017,Browaeys2020,Ebadi2020}, and long coherence times
in state-independent optical traps~\cite{Madjarov2019,Norcia2019,Young2020} --
for the first time to enable new directions in quantum communication and
distributed quantum computing. Moreover, we have demonstrated how this platform
can offer dramatic improvements in entanglement generation rates over long
distances by time-multiplexing across an array of atoms within each entanglement
generation attempt.

We show that entanglement generation rates with $N \approx 100$ atoms across $\gtrsim
100\,\t{km}$-links compare favorably with conservative estimates of the atoms'
coherence time. We further demonstrate that multiplexed repeater-based networks
with $2^{(m = 4)}$ links and $N \approx 100$ atoms at each node can generate
entanglement over $\approx 1500\,\t{km}$. Additionally, we show that our system
is well-suited for entanglement
purification~\cite{Dur2003,Bennett1996a,Kalb2017} and can achieve a purified
network range to $\approx 1100\,\t{km}$, providing a promising architecture for
a transcontinental quantum network. This network architecture is also compatible
with heterogeneous hardware, and may be combined with microwave-to-optical
transduction~\cite{Covey2019c,Lauk2020} to provide a robust network between
superconducting quantum processors~\cite{Arute2019}. Finally, we consider the
prospects for generating larger numbers of Bell pairs for more advanced
protocols such as distributing logically-encoded or other many-body states
relevant for quantum computing and metrology. We find that $26$ Bell pairs can
be generated over a metropolitan link of $50\,\t{km}$.

More generally, the confluence of the associated research thrusts -- Rydberg
atoms arrays~\cite{Saffman2010,Browaeys2020}, cavity QED with strong atom-photon
coupling~\cite{McKeever2003,Birnbaum2005,Tiecke2014}, and atom-array optical
clocks~\cite{Madjarov2019,Norcia2019,Young2020} -- into one platform will enable
new methods to engineer, measure, and distribute many-body entangled states with
single-qubit control. For example, the optical cavity can mediate non-demolition
measurements~\cite{Boca2004,Kalb2015} that could augment the Rydberg-based
quantum computing platform. Conversely, Rydberg-mediated interactions and
single-atom control may help to enhance and distribute spin-squeezed states of
optical clock qubits generated via the
cavity~\cite{Polzik2016,Pezze2018,Pedrozo2021}. Finally, the marriage of
short-ranged (Rydberg-mediated) and infinite-ranged (cavity-mediated)
interactions combined with the possibility of atom-selective control and readout
will enable new opportunities for the study of quantum many-body phenomena such
as the simulation of magnetism~\cite{Davis2019} and chaotic
dynamics~\cite{Choi2021} in regimes not readily accessible to classical
computers.

\section*{Acknowledgments}
We thank Michael Bishof and Liang Jiang for stimulating discussions and Johannes
Borregaard for carefully reading this manuscript. We acknowledge funding from
the NSF QLCI for Hybrid Quantum Architectures and Networks (NSF award 2016136).
\vspace{5mm}

\setcounter{section}{0}
\twocolumngrid

\renewcommand\thefigure{S\arabic{figure}}
\renewcommand\thetable{S\arabic{table}}
\setcounter{figure}{0}  

\renewcommand\appendixname{APPENDIX}
\appendix
\renewcommand\thesection{\Alph{section}}
\renewcommand\thesubsection{\arabic{subsection}}


\section{Quantum repeater and purification protocols}
\label{Sec:RepPur}
\begin{figure}[t!]
    \centering
    \includegraphics[width=7cm]{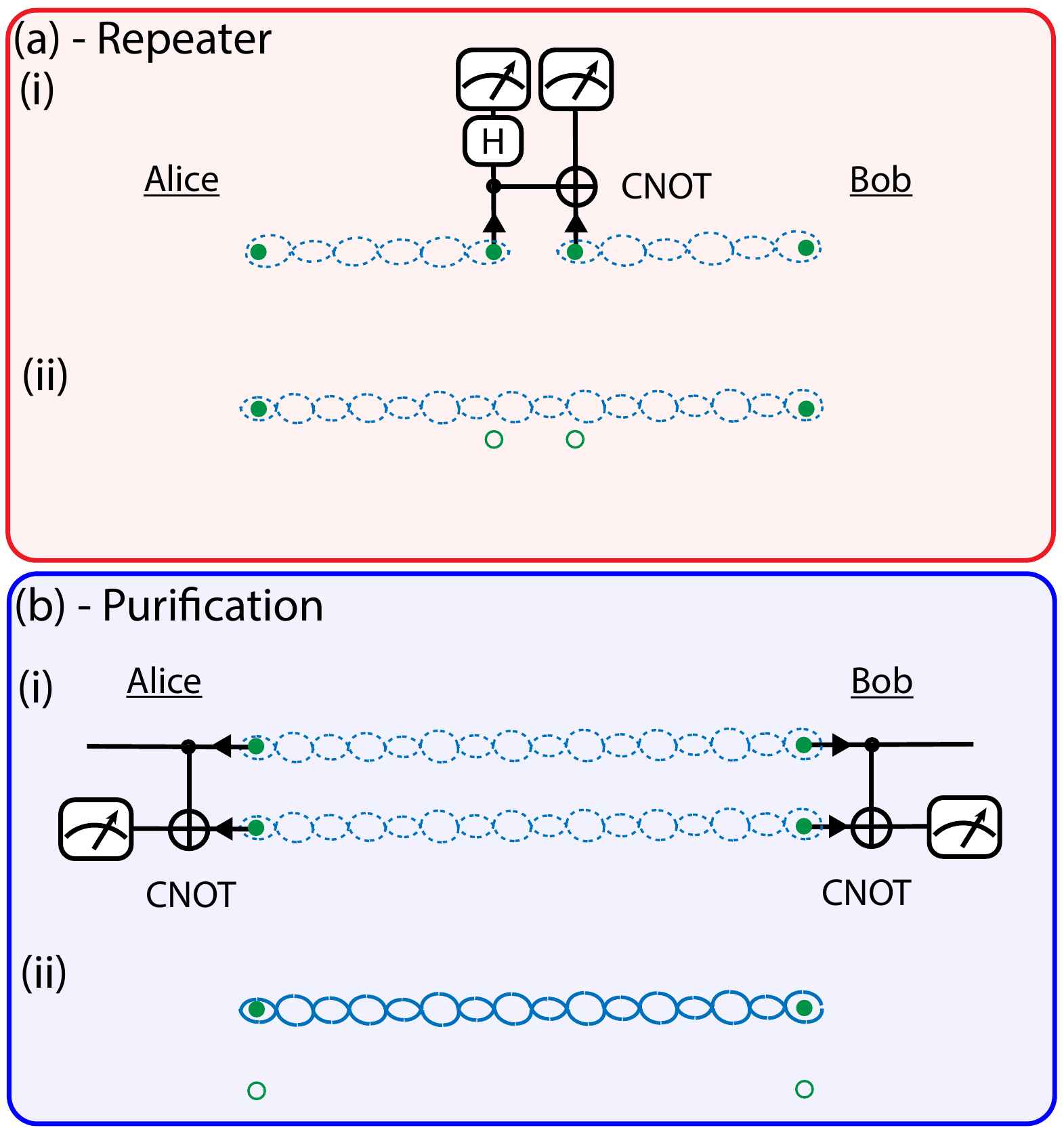}
    \caption{
        \textbf{Schematic overview of quantum repeater and purification
        protocols}. See text for details.
    }
    \label{FigS1}
\end{figure}

The repeater protocol~\cite{Duan2001,Pirandola2019} is based on creating two
Bell pairs, where end-users Alice and Bob each have half of one pair and the
intermediate node has half of both. Then, the combination of a deterministic
two-qubit controlled-NOT gate (CNOT), a single-qubit Hadamard gate, and the
qubit measurements swaps the entanglement out of the two qubits at the
intermediate node and leaves Alice and Bob's halves entangled in a Bell pair
with no quantum information remaining at the intermediate node [see
Fig.~\ref{FigS1}(a)].

The purification protocol~\cite{Dur2003,Bennett1996a,Kalb2017} is based on
creating two Bell pairs, where end-users Alice and Bob each have half of both
pairs. A CNOT gate and a single-qubit measurement at both nodes leaves only one
Bell pair between Alice and Bob that has higher fidelity than either initial
pair. No quantum information remains in the other qubit pair [see
Fig.~\ref{FigS1}(b)]. This protocol could be extended to the case of
intermediate nodes and could be combined with the repeater protocol. We note
that all necessary inner-node single- and two-qubit operations, and measurements
for these protocols have been demonstrated in atomic
arrays~\cite{Levine2019,Graham2019,Madjarov2020}.

\section{Atomic and cavity QED parameters}
\label{Sec:Atomic}

\subsection{Notes on the Yb telecom-band transitions}
We begin this section by compiling a list of references for the Yb
$\PotoD{J}{J'}$
transitions~\cite{Bowers1996,Porsev1999,Loftus2002,Dzuba2010,Guo2010,Cho2012,Beloy2012,Lee2015,Antypas2019}.
In the literature, there appears to be universal agreement that the decay rate
from ${}^3 \t{D}_2$ to ${}^3 \t{P}_1$ (the transition of interest in this work)
is $\Gamma_{\DtoP{2}{1}} = 2 \times 10^6\,\t{s}^{-1}$ and the decay rate from
${}^3 \t{D}_2$ to ${}^3 \t{P}_2$ is $\Gamma_{\DtoP{2}{2}} = 2 \times
10^5\,\t{s}^{-1}$. This corresponds to a branching ratio of the desired decay
path of $0.87$.

However, there is disagreement about the decay rates of ${}^3 \t{D}_1$ to ${}^3
\t{P}_J$. In particular, the literature is split between $\{ \Gamma_\DtoP{1}{0},
\Gamma_\DtoP{1}{1}, \Gamma_\DtoP{1}{2} \} = \{ 200, 100, 3 \} \times
10^4$~\cite{Bowers1996,Porsev1999} and $\{ 200, 10, 3 \} \times
10^4\,\t{s}^{-1}$~\cite{Loftus2002}. We believe that Ref.~\cite{Loftus2002} --
which came after Refs.~\cite{Bowers1996,Porsev1999} -- introduced an error that
has since propagated in the community.
References~\cite{Cho2012,Lee2015,Covey2019b,Covey2019c} have propagated this
error, though it has not affected their arguments or conclusions, while
Refs.~\cite{Antypas2019} and others use the correct values.

\subsection{Cavity QED parameters}
We consider a cavity characterized by two parameters: the radius of curvature $R
= 5 \,\t{mm}$ of its two mirrors and the length $\ell = 9.75 \,\t{mm}$ between
them. For these parameters, the cavity is near-concentric, and satisfies the
stability condition $0 \leq \mathcal{G}^2 \leq 1$, where $\mathcal{G} = 1 - \ell
/ R$ is the cavity stability parameter. We also characterize the principal mode
of the cavity by its waist $w_0$, Rayleigh range $z_0$, and volume $V_m$ using
\begin{align}
    w_0
        &= \left[
            \left( \frac{\lambda_\telecom \ell}{2 \pi} \right)^2
            \left( \frac{1 + \mathcal{G}}{1 - \mathcal{G}} \right)
        \right]^{1/4}
    ,\\
    z_0
        &= \frac{\pi w_0^2}{\lambda_\telecom}
    ,\\
    V_m
        &= \frac{\pi}{4} w_0^2 \ell
,\end{align}
where $\lambda_\telecom$ is the wavelength of the targeted telecom transition,
$1480 \,\t{nm}$. We also assume the cavity to have intrinsic finesse $F_\t{int}
= 10^5$, transmission linewidth $\kappa_\t{int} = 2 \pi \times c / 2 \ell
F_\t{int} \approx 2 \pi \times 154 \,\t{kHz}$, and free spectral range $\t{FSR}
= c / 2 \ell \approx 15.4 \,\t{GHz}$. For a chosen extrinsic finesse $F_\t{ext}
= 5 \times 10^4$ ($\kappa_\t{ext} \approx 2 \pi \times 307 \,\t{kHz}$), this
gives a photon collection efficiency of $\eta_\t{coll} = 1 - F_\t{ext} /
F_\t{int} = 0.5$.  Now we consider the atom-cavity interaction parameters
essential to the proposed scheme. The electric dipole matrix element $D$ for our
chosen transition is
\begin{equation}
    \begin{split}
    D = \Bigg[
        & \frac{3 \pi \varepsilon_0 \hbar c^3}{\omega_\telecom^3}
        \Gamma_\telecom
        \\
        & \phantom{\times} 
        \begin{Bmatrix} J & J' & 1 \\ F' & F & I \end{Bmatrix}
        (2 F' + 1) (2 J' + 1)
    \Bigg]^{1/2}
    ,\end{split}
\end{equation}
where $\omega_\telecom \approx 2 \pi \times 202 \,\t{THz}$, $\Gamma_\telecom = 2
\pi \times 318 \,\t{kHz}$ and the term in braces is the Wigner 6-$j$ symbol,
giving $D \approx 1.96 \times 10^{-29} \,\t{C\,m}$. Using this, the coherent
coupling to the cavity mode is
\begin{equation}
    g_{34}
        = \frac{D}{\hbar}
            \left[
        \frac{\hbar \omega_\telecom}{2 \varepsilon_0 V_m}
        \right]^{1/2}
    \approx 2 \pi \times 1.53 \,\t{MHz}
,\end{equation}
which gives cooperativity $C = g_{34}^2 / \kappa \Gamma_\telecom \approx 16.0$
with $\kappa = \kappa_\t{int} + \kappa_\t{ext}$. Then the probability of
emitting a telecom photon into the cavity mode is $P_\t{cavity} = C / (C + 1)
\approx 0.941$, and hence the probability of extracting this photon for use in
our scheme is $\eta_\t{extract} = P_\t{cavity} \eta_\t{coll} \approx 0.471$.

\section{Four-wave mixing}\label{Sec:FWM}

\begin{figure}[t!]
    \centering
    \includegraphics[width=2.25in]{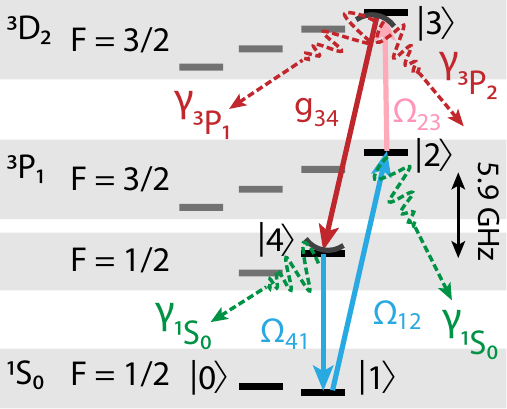}
    \caption{
        \textbf{Relevant levels of four-wave mixing}. States $\ket{0}$ to
        $\ket{4}$ were considered for the simulation to estimate the success
        probability and fidelity. During the excitation cycle, the population in
        the levels $\ket{2}$ and $\ket{4}$ decay to state ${}^1 \t{S}_0$ with a
        decay rate $\gamma_{{}^1 \t{S}_0} = 2 \pi \times 182\,\t{kHz}$, and
        population in state $\ket{3}$ decays to state ${}^3 \t{P}_1$ and ${}^3
        \t{P}_2$ with decay rate $\gamma_{{}^3 \t{P}_1} = 2 \pi \times
        318\,\t{kHz}$ and $\gamma_{{}^3 \t{P}_2} = 2 \pi \times 48\,\t{kHz}$,
        respectively. For simulation purposes, all decays are assumed to
        accumulate in a dump level that does not contribute to the coherent
        evolution.
    }
    \label{FigS2}
\end{figure}

\subsection{Numerical model and results}
The atom-telecom photon entanglement generation protocol is similar to the
four-level scheme previously shown for rubidium and cesium atoms coupled to
nanophotonic cavities \cite{Menon2020}. The protocol starts with initializing
atom in the superposition state $(\ket{0}_a + \ket{1}_a) / \sqrt{2}$. This is
followed by a pulse sequence that takes the atom through states $\ket{1}
\rightarrow \ket{4}$ before returning back to the initial state $\ket{1}$.
First, pulse $\Omega_{12}$ transfers population from state $\ket{1}$ to
$\ket{2}$. Then the population is excited to state $\ket{3}$ by light field
$\Omega_{23}$, which is always on. The population that reaches the state
$\ket{3}$ is preferentially transferred to state $\ket{4}$ via the emission of a
telecom photon into the cavity, which is resonant with the $\ket{3}
\leftrightarrow \ket{4}$ transition. A second pulse, $\Omega_{41}$, then
transfers the population in the state $\ket{4}$ back to state $\ket{1}$. The
spontaneous decay from excited states (see Fig.~\ref{FigS2}) limit the coherent
completion of this cycle and leads to infidelities. Here we define the fidelity
as the probability of finding the atom in the qubit state after the round-trip
through states $\ket{1} \rightarrow \ket{4}$, given the heralding of the telecom
photon.

The requirement of heralding makes this scheme robust to any atomic decays
preceding the photon emission into the cavity and limits the infidelities to
decays from the state $\ket{4}$. The optimum parameters for the given pulse
sequence are extracted using a two-step optimization process \cite{Menon2020}.
The first step optimizes the Rabi frequencies $\Omega_{12},\, \Omega_{23}$ and
the pulse width of $\Omega_{12}$ to maximize the population transfer to the
state $\ket{4}$ and the second step optimizes the timing, pulse width, and Rabi
frequency of $\Omega_{41}$. In both the schemes below, the success probability
accounts for the probability $P_\ket{1}$ for the initial population in $\ket{1}$
to emit a telecom photon and return to $\ket{1}$, as well as the probability for
the emitted photon to couple to the external coupling mode of the cavity; i.e.
\begin{equation}
    \text{Success probability}
        = \frac{\kappa_\t{ext}}{\kappa_\t{int}} P_\ket{1}
\end{equation}

\subsubsection{Resonant case}
In the first case, which we call the ``resonant case,'' we have the cavity on
resonant with the $\ket{3} \leftrightarrow \ket{4}$ transition. In this case the
corresponding Hamiltonian in an appropriately chosen rotating frame is
\begin{equation}
    \begin{split}
        \hat{H}
            &= \Omega_{12}(t) \ket{1} \bra{2}
            + \Omega_{12} \ket{2} \bra{3}
            + g_{34} \hat{a} \ket{3} \bra{4}
            \\ & \phantom{=}
            + \Omega_{41}(t) \ket{4} \bra{1} + \text{H.c.}
    \end{split}
\end{equation}

In this resonant excitation scheme, the population transfer to $\ket{4}$ occurs
over a time scale that is inversely proportional to atom-cavity coupling
$g_{34}$, and for efficient completion, the second pulse has to be timed to
match. The earlier coherent transfer spend a longer time in $\ket{4}$ leading to
spontaneous decay. To minimize the contribution to infidelity, we transfer the
population from $\ket{4}$ at earlier times, trading fidelity gains for reduced
efficiency, due to incomplete population transfer. Here we achieve this by
applying $\Omega_{41}$ earlier than what is optimal for the complete population
transfer shown in Fig. \ref{Fig3}(c). The increase in fidelity and corresponding
reduction in the efficiency are shown in Fig.\ref{Fig3}(d). Fixed Gaussian
pulses with full widths at half maximum $116\,\t{ns}$ and $58\,\t{ns}$ were used
for $\Omega_{12}$ and $\Omega_{41}$, respectively. Here, the achieved fidelities
were conditioned on heralding entanglement using the photons that were emitted
until the coherent transfer back to the initial qubit state by $\Omega_{41}$.
Detection of photons emitted from the cavity after the completion of
$\Omega_{41}$ leads to additional infidelities.\\ 

\begin{figure}[t!]
    \centering
    \includegraphics[width=8.6cm]{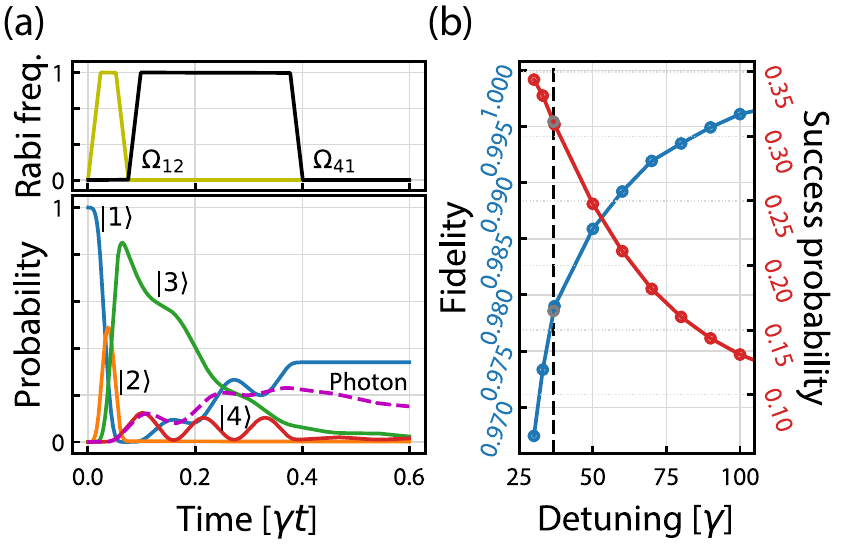}
    \caption{
        \textbf{Detuned four-wave mixing scheme}. (a) Time evolution of the
        coherent population in the states $\ket{1}$ to $\ket{4}$ and the photon
        emitted from the cavity through external coupling parameter
        $\kappa_\t{ext}$. The top half of the plot shows pulses $\Omega_{12}$
        and $\Omega_{41}$. In the simulation, the first pulse is assumed to have
        a constant length of $65\,\t{ns}$ and the length of the second pulse is
        optimized for each value of the detuning. Here, both pulses have a
        $20\,\t{ns}$ ramp time. (b) Scaling of fidelity and success probability
         with detuning. Larger detuning leads to lower occupation and
        decay from state $\ket{4}$ at the cost of lower coherent population
        transfer back to state $\ket{1}$.
    }
    \label{FigS3}
\end{figure}

\subsubsection{Detuned case}
High-fidelity atom-telecom photon entanglement can also be obtained by using an
off-resonant scheme, where the population transfer to $\ket{4}$ is minimized,
since decay from this state is the dominant error in the heralding protocol. In
this case the Hamiltonian considered is
\begin{equation}
    \begin{split}
        \hat{H}
            &= \Omega_{12}(t) \ket{1} \bra{2}
            + \Omega_{23} \ket{2} \bra{3}
            + g_{34} \hat{a} \ket{3} \bra{4}
            + \Omega_{41}(t) \ket{4} \bra{1}
            \\ & \phantom{=}
            + \delta \ket{4} \bra{4} + \text{H.c.}
    \end{split}
\end{equation}

In this scheme, the optimal fidelities were also found by a two-step
optimization procedure. For a given detuning, the first step maximized the
population transfer to $\ket{3}$ by optimizing the Rabi frequencies
$\Omega_{12},\, \Omega_{23}$ and the pulse width of $\Omega_{12}$, and the
second step optimizes the duration and Rabi frequency of $\Omega_{41}$ to
maximize the population transfer from $\ket{3}$ to $\ket{1}$ through the
two-photon process. Here, we fix the pulse length of $\Omega_{12}$ to
$65\,\t{ns}$ including a linear ramp time of $20\,\t{ns}$. The length of
$\Omega_{41}$ varies from $300\,\t{ns}$ to $500\,\t{ns}$ according to the varied
detuning. Similar to the resonant case we again find that higher fidelities can
be obtained at the cost of lower success probabilities [see
Fig.~\ref{FigS3}(b)]. Incomplete population transfer in both schemes will lead
to some residual population left behind in the states that are coupled to the
cavity, which can lead to photon emission even after the end of the pulse
sequence. Detection of these photons will add to infidelity. Overall success
probabilities were found to be greater for the resonant scheme that is used in
the main text for our calculations.

\subsection{Phase matching considerations}
We consider the importance of phase matching and momentum conservation of the
four light fields that have overlapping amplitude during our four-wave scheme.
We perform a qualitative estimate based on classical four-wave mixing analysis
in which an outgoing wave is produced by the interaction of three incoming waves
with a nonlinear medium~\cite{Ender1982}. The outgoing field intensity is
proportional to a phase-matching factor whose argument is $\xi = \Delta k \times
L$, where $\Delta k = |\vec{k}_{12} + \vec{k}_{23} - \vec{k}_{34} -
\vec{k}_{41}|$ and $L$ is the effective overlap length of the four fields which
in practice is determined by their size or the size of the medium (whichever is
smaller). The phase matching factor is equal to one when $\xi = 0$ and decreases
for $\xi \gg 0$.

For the beam configuration shown in Fig.~\ref{Fig3}(b) assuming a $180^\circ$
angle between $\vec{k}_{12}$ and $\vec{k}_{41}$ and a $45^\circ$ angle between
$\vec{k}_{34}$ and $\vec{k}_{23}$, we estimate that $\Delta k \approx 2 \pi /
(1500 \sqrt{2})\,\t{nm}$. The relevant length scale of the single-atom case
should be the size of the atomic wavefunction in the optical trap, which we
assume is $L_\t{single} \approx 50\,\t{nm}$. Hence, we estimate that $\xi < 1$
for the case we consider here, so phase matching of the four light fields is not
crucial. We therefore neglect it in our analysis, but choose a beam geometry to
minimize $\Delta k$.
  
For an atomic ensemble or a solid-state spin ensemble, this factor would be much
higher. Assuming $L = 10~ \mu$m with the same beam geometry, $\xi \gg 10$.
Hence, phase matching is often crucial in ensemble and crystalline environments.

\section{Entanglement distribution calculations}
\label{Sec:Dist}
We start by considering the rate $\Gamma_\t{link}$ at which entanglement between
two adjacent network nodes can be attempted. This rate comprises all components
shown in Fig.~\ref{Fig2}. The time to cool and initialize all atoms at the nodes
(shown in blue in Fig.~\ref{Fig2}), performed globally and in parallel over the
arrays of atoms at both nodes, is $1 / \Gamma_\t{init} = 1 / 10\,\t{kHz}$. This
is based on the maximum scattering rate from the $^3$P$_1$
($\Gamma_{^3\text{P}_1}\approx90$ kHz) and an assumption about the number of
photons required for cooling and optical pumping. The total qubit pulse time
comprising globally applied $\pi / 2$- and $\pi$-pulses (all time windows shown
in green in Fig.~\ref{Fig2}) is $1 / \Gamma_{\pi / 2} + 1 / \Gamma_\pi = 3 /
\Gamma_{\pi / 2} = 3 / 100\,\t{kHz}$, based on an assumed Rabi frequency of 50
kHz via stimulated Raman pulses between the nuclear spin states. The total
four-wave-mixing time (shown in orange in Fig.~\ref{Fig2}) is $2 N /
\Gamma_\t{FWM} = 2 N / 200\,\t{kHz}$ for all $N$ atoms. The four-wave-mixing
rate is determined by the time between when the sequence begins and when the
photon leaves the cavity with high probability [see Fig.~\ref{Fig3}(c)].
Finally, the time to transmit classical signals through fibers and to herald
entanglement (shown in purple in Fig.~\ref{Fig2}) is $1 / \Gamma_\t{comm}(L) = c
/ 2 L \approx 10^8\,\t{m\,s$^{-1}$} / L$.

\begin{table*}
    \centering
    \caption{
        \textbf{Rates for constituent steps in the single-link multiplexing
        protocol}. Exact values for rates composing the total rate at which
        entanglement between two adjacent network nodes $\Gamma_\t{link}$ can be
        attempted (Eq.~\ref{Eq:Ymux}) in terms of the per-node atom number $N$
        and distance in fiber between nodes $L$. The listed steps correspond
        chronologically to the colored time windows shown in Fig.~\ref{Fig2}(a)
        and (b).
    }
    \label{Tab1}
    \begin{tabular}{lllll}
        \hline\hline
        Step & Symbol & Description & Global? & Rate [kHz]
        \\
        \hline
        1 & $\Gamma_\t{init}$ & Optical pumping and cooling & Global & $10$
        \\
        2 & $\Gamma_{\pi / 2}$ & $\pi / 2$-pulse & Global & $100$
        \\
        3 & $\Gamma_\t{FWM}$ & FWM protocol & One-by-one & $200 / N$
        \\
        4 & $\Gamma_\pi$ & $\pi$-pulse & Global & $50$
        \\
        5 & $\Gamma_\t{FWM}$ & FWM protocol & Global (same order) & $200 / N$
        \\
        6 & $\Gamma_\t{comm}(L)$ & Heralded entanglement
            & Global (atom-unique time stamp) & $c / 2 L$
        \\
        \hline\hline
    \end{tabular}
\end{table*}

Using these quantities, $\Gamma_\t{link}$ is
\begin{equation}
    \label{Eq:Ylink}
    \Gamma_\t{link} (L, N)
        = \left[
            \frac{1}{\Gamma_\t{init}}
            + \frac{3}{\Gamma_{\pi / 2}}
            + \frac{2 N}{\Gamma_\t{FWM}}
            + \frac{1}{\Gamma_\t{comm} (L)}
        \right]^{-1}
.\end{equation}
The exact values used for these rates are summarized in Table~\ref{Tab1}.

The probability $p$ of successfully creating a single Bell pair between any
given two atoms at adjacent nodes similarly comprises several components;
\begin{equation}
    p (L)
        = \left( \frac{1}{2} \right)^2
            \eta_\t{FWM}^2
            \, \eta_\t{fiber}^2
            \, \eta_\t{det}^2
            \, \eta_\t{att} (L)
.\end{equation}
Here, $\eta_\t{FWM} \approx 0.364$ is the total success probability of the
four-wave-mixing scheme under the condition shown in Fig.~\ref{Fig3}(d);
$\eta_\t{fiber} = \eta_\t{det} = 0.9$ are the efficiencies at which photons may
be collected by their respective fibers and subsequently detected; and
$\eta_\t{att}(L) = \exp(-L / \lambda_\t{att})$ is the attenuation of the telecom
photons ($\lambda_\t{att} = 20.7 \,\t{km}$ at
$1480\,\t{nm}$~\cite{Corning2020}). The two leading factors of $1 / 2$ are due
to the overlap between the Bell-state and computational bases and an assumed
complete loss of photon polarization in the long-distance fibers. It follows
that the total probability $P_\t{mux}$ of creating at least $B$ Bell pairs
between adjacent nodes through multiplexing is
\begin{equation}
    \label{Eq:Pmux}
    P_\t{mux} (L, N, B)
        = \sum\limits_{k=B}^N \binom{N}{k} p^k (1 - p)^{N - k}
\end{equation}
for $N \geq B$ and zero otherwise.

To calculate the rate $\Gamma_\t{mux}$ at which these $B$ or more Bell pairs can
be formed between atoms at adjacent network nodes, we consider a total number of
times $M$ that the entire procedure is attempted. While $M$ is in principle
unbounded, it is realistic to choose $M$ such that the mean number of
successful attempts $M P_\t{mux} (L, N, B)$ to create $\geq B$ Bell pairs is
one, and hence the average success rate is that at which these $M$ attempts can
be performed,
\begin{equation}
    \label{Eq:Ymux}
    \Gamma_\t{mux} (L, N, B)
        = \Gamma_\t{link} (L, N) \times P_\t{mux} (L, N, B)
.\end{equation}

\begin{figure}[t!]
    \centering
    \includegraphics[width=8.6cm]{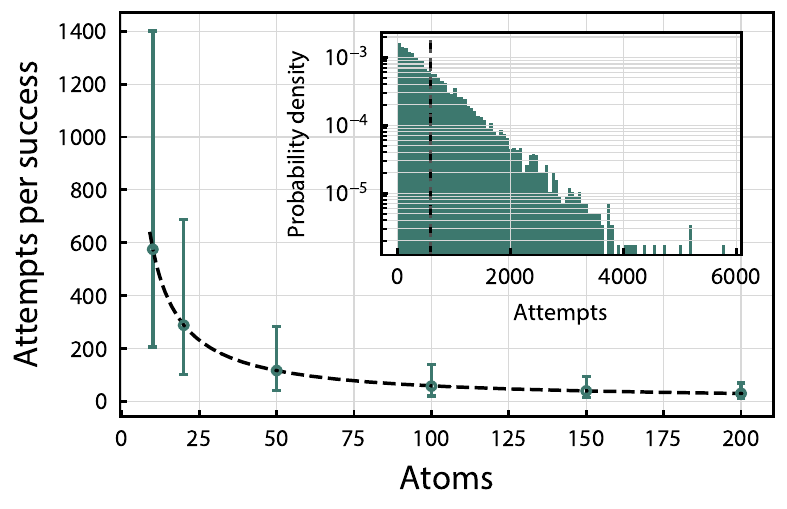}
    \caption{
        \textbf{Comparison of simulation with Eq.~\ref{Eq:Pmux}}. The simulated
        mean number of attempts $M$ required for the formation of one Bell pair
        across a single link for various $N$ at $L = 100\,\t{km}$ (green
        circles), averaged over $10,000$ trials, are compared in the main plot
        with the expected analytically derived result $M = 1 / P_\t{mux}$ (black
        dashed line). The error bars show the one-sided RMS deviation from the
        mean. The inset shows the distribution over values of $M$ associated
        with the $N = 10$ data point along with the mean of the distribution
        (gray dotted line) and analytical result (black dashed line).
    }
    \label{FigS4}
\end{figure}

Generalizing to the network-level procedure, the two-group structure [see
Sec.~\ref{sec:network level}] requires an extra consideration. The proposed
protocol requires that entanglements in Group 1 complete before those in Group 2
can be attempted, which precludes the derivation of an analytical formula to
describe the expected rates; hence we turn to numerical simulation to calculate
the results shown in Figs.~\ref{Fig5} and \ref{Fig6}. We use a simple simulation
scheme based on stochastically sampling the probability distribution over $M$
attempts required for the formation of Bell pairs across each network link in
accordance with Eq.~\ref{Eq:Pmux}, which gives $M = 1 / P_\t{mux}$ for one
success across a link for given $N$, $L$, and $B$. For our simulations, $M$ -- which follows a geometric distribution -- is
sampled by counting the number of random events required for a single success,
which occurs with probability $P_\t{mux}$. The resulting averages of these
counts over $10,000$ trials are in good agreement with the expected value
obtained using Eq.~\ref{Eq:Pmux}, shown in Fig.~\ref{FigS4}. The time taken for
each linking attempt is then $M / \Gamma_\t{link}$, and the mean of a set of $Q$
such trials can be inverted to find the average entanglement rate. It was found
that this scheme could be used for as few as $Q = 1000$ trials to
faithfully reproduce Fig.~\ref{Fig4}.

At the network level, the results shown in Figs.~\ref{Fig5} and \ref{Fig6}(a)
were calculated following the two-group protocol as described. The single-link
linking time $M / \Gamma_\t{link}$ was singly sampled for each of the
$2^{m - 1}$ network links in Group 1, from which the maximum was selected. This
sampling was repeated for $2^{m - 1}$ links in Group 2 (for $N - 1$ atoms at
each node), and the two maxima were added to find the total time required for
the network. The mean of $Q = 5000$ such trials was then inverted to calculate
the average rate for each value of $m$, $N$, and $L$ shown. For the multi-Bell
case shown in Fig.~\ref{Fig6}(b), the single single-link procedure described in
the previous paragraph was repeated for $B$ Bell pairs following the ``ladder''
scheme [see Sec.~\ref{sec:multi-bell}], the total time for $B$ linking attempts
averaged over $Q = 5000$ trials, and the average inverted for each value of $N$
and $B$ shown.

\bibliographystyle{h-physrev}
\bibliography{library}

\begin{thebibliography}{10}

\bibitem{Cirac1997}
J.~I. Cirac, P.~Zoller, H.~J. Kimble, and H.~Mabuchi,
\newblock Phys. Rev. Lett. {\bf 78}, 3221 (1997).

\bibitem{Wehner2018}
S.~Wehner, D.~Elkouss, and R.~Hanson,
\newblock Science {\bf 362}, eaam9288 (2018).

\bibitem{Kimble2008}
H.~J. Kimble,
\newblock Nature {\bf 453}, 1023 (2008).

\bibitem{Pirandola2019}
S.~Pirandola {\em et~al.},
\newblock Adv. Opt. Photonics {\bf 12}, 1012 (2020).

\bibitem{Jiang2007}
L.~Jiang, J.~M. Taylor, A.~S. S{\o}rensen, and M.~D. Lukin,
\newblock Phys. Rev. A {\bf 76}, 062323 (2007).

\bibitem{Komar2014}
P.~K{\'{o}}m{\'{a}}r {\em et~al.},
\newblock Nat. Phys. {\bf 10}, 582 (2014).

\bibitem{Young2020}
A.~W. Young {\em et~al.},
\newblock Nature {\bf 588}, 408 (2020).

\bibitem{Ebadi2020}
S.~Ebadi {\em et~al.},
\newblock Nature {\bf 595}, 227 (2021).

\bibitem{Saffman2010}
M.~Saffman, T.~G. Walker, and K.~M{\o}lmer,
\newblock Rev. Mod. Phys. {\bf 82}, 2313 (2010).

\bibitem{Levine2019}
H.~Levine {\em et~al.},
\newblock Phys. Rev. Lett. {\bf 123}, 170503 (2019).

\bibitem{Graham2019}
T.~M. Graham {\em et~al.},
\newblock Phys. Rev. Lett. {\bf 123}, 230501 (2019).

\bibitem{Uphoff2016}
M.~Uphoff, M.~Brekenfeld, G.~Rempe, and S.~Ritter,
\newblock Appl. Phys. B {\bf 122}, 46 (2016).

\bibitem{Covey2019b}
J.~P. Covey {\em et~al.},
\newblock Phys. Rev. Appl. {\bf 11}, 034044 (2019).

\bibitem{Menon2020}
S.~G. Menon, K.~Singh, J.~Borregaard, and H.~Bernien,
\newblock New J. Phys. {\bf 22}, 073033 (2020).

\bibitem{Reiserer2015}
A.~Reiserer and G.~Rempe,
\newblock Rev. Mod. Phys. {\bf 87}, 1379 (2015).

\bibitem{Ritter2012}
S.~Ritter {\em et~al.},
\newblock Nature {\bf 484}, 195 (2012).

\bibitem{Hofmann2012}
J.~Hofmann {\em et~al.},
\newblock Science {\bf 337}, 72 (2012).

\bibitem{Samutpraphoot2020}
P.~Samutpraphoot {\em et~al.},
\newblock Phys. Rev. Lett. {\bf 124}, 063602 (2020).

\bibitem{Langenfeld2021}
S.~Langenfeld {\em et~al.},
\newblock Phys. Rev. Lett. {\bf 126}, 130502 (2021).

\bibitem{Daiss2021}
S.~Daiss {\em et~al.},
\newblock Science {\bf 371}, 614 (2021).

\bibitem{Dordevic2021}
T.~Dordevi{\'{c}} {\em et~al.},
\newblock arXiv Prepr. {\bf 2105.06485} (2021).

\bibitem{Duan2001}
L.-M. Duan, M.~D. Lukin, J.~I. Cirac, and P.~Zoller,
\newblock Nature {\bf 414}, 413 (2001).

\bibitem{Pfaff2013}
W.~Pfaff {\em et~al.},
\newblock Nat. Phys. {\bf 9}, 29 (2013).

\bibitem{Graham2013}
T.~M. Graham, J.~T. Barreiro, M.~Mohseni, and P.~G. Kwiat,
\newblock Phys. Rev. Lett. {\bf 110}, 060404 (2013).

\bibitem{Sinclair2014}
N.~Sinclair {\em et~al.},
\newblock Phys. Rev. Lett. {\bf 113}, 053603 (2014).

\bibitem{Kaneda2015}
F.~Kaneda {\em et~al.},
\newblock Optica {\bf 2}, 1010 (2015).

\bibitem{Zhong2017}
T.~Zhong {\em et~al.},
\newblock Science {\bf 357}, 1392 (2017).

\bibitem{Wengerowsky2018}
S.~Wengerowsky, S.~K. Joshi, F.~Steinlechner, H.~H{\"{u}}bel, and R.~Ursin,
\newblock Nature {\bf 564}, 225 (2018).

\bibitem{Corning2020}
Corning SMF-28 Ultra Opt. Fiber , https://www.corning.com/.

\bibitem{Dur2003}
W.~D{\"{u}}r and H.-J. Briegel,
\newblock Phys. Rev. Lett. {\bf 90}, 067901 (2003).

\bibitem{Bennett1996a}
C.~H. Bennett {\em et~al.},
\newblock Phys. Rev. Lett. {\bf 76}, 722 (1996).

\bibitem{Kalb2017}
N.~Kalb {\em et~al.},
\newblock Science {\bf 356}, 928 (2017).

\bibitem{Barrett2005}
S.~D. Barrett and P.~Kok,
\newblock Phys. Rev. A {\bf 71}, 060310 (2005).

\bibitem{Bernien2013}
H.~Bernien {\em et~al.},
\newblock Nature {\bf 497}, 86 (2013).

\bibitem{Endres2016}
M.~Endres {\em et~al.},
\newblock Science {\bf 354}, 1024 (2016).

\bibitem{Barredo2016}
D.~Barredo, S.~de~L{\'{e}}s{\'{e}}leuc, V.~Lienhard, T.~Lahaye, and
  A.~Browaeys,
\newblock Science {\bf 354}, 1021 (2016).

\bibitem{Cooper2018}
A.~Cooper {\em et~al.},
\newblock Phys. Rev. X {\bf 8}, 041055 (2018).

\bibitem{Norcia2018b}
M.~A. Norcia, A.~W. Young, and A.~M. Kaufman,
\newblock Phys. Rev. X {\bf 8}, 041054 (2018).

\bibitem{Saskin2019}
S.~Saskin, J.~T. Wilson, B.~Grinkemeyer, and J.~D. Thompson,
\newblock Phys. Rev. Lett. {\bf 122}, 143002 (2019).

\bibitem{Casabone2013}
B.~Casabone {\em et~al.},
\newblock Phys. Rev. Lett. {\bf 111}, 100505 (2013).

\bibitem{Nguyen2018}
C.~H. Nguyen, A.~N. Utama, N.~Lewty, and C.~Kurtsiefer,
\newblock Phys. Rev. A {\bf 98}, 063833 (2018).

\bibitem{Davis2019}
E.~J. Davis, G.~Bentsen, L.~Homeier, T.~Li, and M.~H. Schleier-Smith,
\newblock Phys. Rev. Lett. {\bf 122}, 010405 (2019).

\bibitem{Zeiher2020}
J.~Zeiher, J.~Wolf, J.~A. Isaacs, J.~Kohler, and D.~M. Stamper-Kurn,
\newblock arXiv Prepr. {\bf 2012.01280} (2020).

\bibitem{Covey2019c}
J.~P. Covey, A.~Sipahigil, and M.~Saffman,
\newblock Phys. Rev. A {\bf 100}, 012307 (2019).

\bibitem{Covey2019}
J.~P. Covey, I.~S. Madjarov, A.~Cooper, and M.~Endres,
\newblock Phys. Rev. Lett. {\bf 122}, 173201 (2019).

\bibitem{Norcia2019}
M.~A. Norcia {\em et~al.},
\newblock Science {\bf 366}, 93 (2019).

\bibitem{Madjarov2019}
I.~S. Madjarov {\em et~al.},
\newblock Phys. Rev. X {\bf 9}, 041052 (2019).

\bibitem{Madjarov2020}
I.~S. Madjarov {\em et~al.},
\newblock Nat. Phys. {\bf 16}, 857 (2020).

\bibitem{Wilson2019}
J.~Wilson {\em et~al.},
\newblock arXiv Prepr. {\bf 1912.08754} (2019).

\bibitem{Polzik2016}
E.~S. Polzik and J.~Ye,
\newblock Phys. Rev. A {\bf 93}, 021404 (2016).

\bibitem{Bernien2017}
H.~Bernien {\em et~al.},
\newblock Nature {\bf 551}, 579 (2017).

\bibitem{Choi2021}
J.~Choi {\em et~al.},
\newblock arXiv Prepr. {\bf 2103.03535} (2021).

\bibitem{Browaeys2020}
A.~Browaeys and T.~Lahaye,
\newblock Nat. Phys. {\bf 16}, 132 (2020).

\bibitem{McKeever2003}
J.~McKeever, A.~Boca, A.~D. Boozer, J.~R. Buck, and H.~J. Kimble,
\newblock Nature {\bf 425}, 268 (2003).

\bibitem{Birnbaum2005}
K.~M. Birnbaum {\em et~al.},
\newblock Nature {\bf 436}, 87 (2005).

\bibitem{Tiecke2014}
T.~G. Tiecke {\em et~al.},
\newblock Nature {\bf 508}, 241 (2014).

\bibitem{Hunger2010}
D.~Hunger {\em et~al.},
\newblock New J. Phys. {\bf 12}, 065038 (2010).

\bibitem{Haas2014}
F.~Haas, J.~Volz, R.~Gehr, J.~Reichel, and J.~Esteve,
\newblock Science {\bf 344}, 180 (2014).

\bibitem{Brekenfeld2020}
M.~Brekenfeld, D.~Niemietz, J.~D. Christesen, and G.~Rempe,
\newblock Nat. Phys. {\bf 16}, 647 (2020).

\bibitem{Sedlacek2016}
J.~A. Sedlacek {\em et~al.},
\newblock Phys. Rev. Lett. {\bf 116}, 133201 (2016).

\bibitem{Thiele2015}
T.~Thiele {\em et~al.},
\newblock Phys. Rev. A {\bf 92}, 063425 (2015).

\bibitem{Teller2021}
M.~Teller {\em et~al.},
\newblock Phys. Rev. Lett. {\bf 126}, 230505 (2021).

\bibitem{Kawasaki2019}
A.~Kawasaki {\em et~al.},
\newblock Phys. Rev. A {\bf 99}, 013437 (2019).

\bibitem{Ye2008}
J.~Ye, H.~J. Kimble, and H.~Katori,
\newblock Science {\bf 320}, 1734 (2008).

\bibitem{Hucul2015}
D.~Hucul {\em et~al.},
\newblock Nat. Phys. {\bf 11}, 37 (2015).

\bibitem{Hansen2015}
B.~Hensen {\em et~al.},
\newblock Nature {\bf 526}, 682 (2015).

\bibitem{Gorshkov2009}
A.~V. Gorshkov {\em et~al.},
\newblock Phys. Rev. Lett. {\bf 102}, 110503 (2009).

\bibitem{Pirandola2017}
S.~Pirandola, R.~Laurenza, C.~Ottaviani, and L.~Banchi,
\newblock Nat. Commun. {\bf 8}, 15043 (2017).

\bibitem{Beugnon2007}
J.~Beugnon {\em et~al.},
\newblock Nat. Phys. {\bf 3}, 696 (2007).

\bibitem{Lengwenus2010}
A.~Lengwenus, J.~Kruse, M.~Schlosser, S.~Tichelmann, and G.~Birkl,
\newblock Phys. Rev. Lett. {\bf 105}, 170502 (2010).

\bibitem{Schymik2020}
K.-N. Schymik {\em et~al.},
\newblock Phys. Rev. A {\bf 102}, 063107 (2020).

\bibitem{Monz2016}
T.~Monz {\em et~al.},
\newblock Science {\bf 351}, 1068 (2016).

\bibitem{Erhard2021}
A.~Erhard {\em et~al.},
\newblock Nature {\bf 589}, 220 (2021).

\bibitem{Fowler2012}
A.~G. Fowler, M.~Mariantoni, J.~M. Martinis, and A.~N. Cleland,
\newblock Phys. Rev. A {\bf 86}, 032324 (2012).

\bibitem{Albert2020}
V.~V. Albert, J.~P. Covey, and J.~Preskill,
\newblock Phys. Rev. X {\bf 10}, 031050 (2020).

\bibitem{Choi2019}
H.~Choi, M.~Pant, S.~Guha, and D.~Englund,
\newblock npj Quantum Inf. {\bf 5}, 104 (2019).

\bibitem{Pezze2018}
L.~Pezz{\`{e}}, A.~Smerzi, M.~K. Oberthaler, R.~Schmied, and P.~Treutlein,
\newblock Rev. Mod. Phys. {\bf 90}, 035005 (2018).

\bibitem{Pedrozo2021}
E.~Pedrozo-Pe{\~{n}}afiel {\em et~al.},
\newblock Nature {\bf 588}, 414 (2020).

\bibitem{Omran2019}
A.~Omran {\em et~al.},
\newblock Science {\bf 365}, 570 (2019).

\bibitem{Song2019}
C.~Song {\em et~al.},
\newblock Science {\bf 365}, 574 (2019).

\bibitem{Marciniak2021}
C.~D. Marciniak {\em et~al.},
\newblock arXiv Prepr. {\bf 2107.01860} (2021).

\bibitem{Lauk2020}
N.~Lauk {\em et~al.},
\newblock Quantum Sci. Technol. {\bf 5}, 020501 (2020).

\bibitem{Arute2019}
F.~Arute {\em et~al.},
\newblock Nature {\bf 574}, 505 (2019).

\bibitem{Boca2004}
A.~Boca {\em et~al.},
\newblock Phys. Rev. Lett. {\bf 93}, 233603 (2004).

\bibitem{Kalb2015}
N.~Kalb, A.~Reiserer, S.~Ritter, and G.~Rempe,
\newblock Phys. Rev. Lett. {\bf 114}, 220501 (2015).

\bibitem{Bowers1996}
C.~J. Bowers {\em et~al.},
\newblock Phys. Rev. A {\bf 53}, 3103 (1996).

\bibitem{Porsev1999}
S.~G. Porsev, Y.~G. Rakhlina, and M.~G. Kozlov,
\newblock Phys. Rev. A {\bf 60}, 2781 (1999).

\bibitem{Loftus2002}
T.~Loftus, J.~R. Bochinski, and T.~W. Mossberg,
\newblock Phys. Rev. A {\bf 66}, 013411 (2002).

\bibitem{Dzuba2010}
V.~A. Dzuba and A.~Derevianko,
\newblock J. Phys. B At. Mol. Opt. Phys. {\bf 43}, 074011 (2010).

\bibitem{Guo2010}
K.~Guo, G.~Wang, and A.~Ye,
\newblock J. Phys. B At. Mol. Opt. Phys. {\bf 43}, 135004 (2010).

\bibitem{Cho2012}
J.~W. Cho {\em et~al.},
\newblock Phys. Rev. A {\bf 85}, 035401 (2012).

\bibitem{Beloy2012}
K.~Beloy {\em et~al.},
\newblock Phys. Rev. A {\bf 86}, 051404 (2012).

\bibitem{Lee2015}
J.~Lee, J.~H. Lee, J.~Noh, and J.~Mun,
\newblock Phys. Rev. A {\bf 91}, 053405 (2015).

\bibitem{Antypas2019}
D.~Antypas {\em et~al.},
\newblock Nat. Phys. {\bf 15}, 120 (2019).

\bibitem{Ender1982}
D.~A. Ender,
\newblock {\em {Doubly-Resonant Two-Photon-Absorption-Induced Four-Wave Mixing
  in Tb(OH)$_3$ and LiTbF$_4$}},
\newblock Ph.d. thesis, Montana State University, 1982.

\end{thebibliography}

\end{document}